%Paper: hep-th/9510100
%From: Ali Chamseddine <chams@itp.phys.ethz.ch>
%Date: Mon, 16 Oct 1995 09:46:21 +0100
%Date (revised): Fri, 17 Nov 1995 09:24:55 +0100

\magnification=\magstep1
\vsize=22.7truecm
\hsize=15.5truecm
\hoffset=.6truecm
\voffset=.8truecm
\parskip=.2truecm

\font\ti=cmbx10 scaled\magstep1
\font\eightrm=cmr8

\def\br{\hfill\break\noindent}

\pageno=0

%
% Titlepage
%
\baselineskip=.5truecm
\footline={\hfill}
%{\hfill ZU-TH- 10/1993}
%\vskip.1truecm
{\hfill ETH/TH/95-27 }
\vskip.2truecm
{\hfill 12 October 1995}
\vskip2.1truecm
\centerline{\ti A Family of Dual N=2 Supergravity Actions in
Ten-Dimensions }
\vskip1.2truecm
\centerline{  A. H. Chamseddine}
\vskip.8truecm
\centerline{ Theoretische Physik}
 \centerline{ ETH-H\"onggerberg}
  \centerline{ CH 8093 Z\"urich }
\centerline{ Switzerland}
\vskip1.2truecm

\centerline{\bf Abstract}
\noindent
We show that non-chiral $N=2$ supergravity in ten-dimensions
admits a family of dual actions where the one-form,
two-form or three-form is replaced by the seven-form,
six-form or five-form respectively. The dual actions and
supersymmetry transformations are given.

\vskip.5truecm

\noindent

\vfill

\eject
\baselineskip=.6truecm
\footline={\hss\eightrm\folio\hss}

%section 1
\centerline{}
\vskip1.1truecm
{\bf\noindent 1. Introduction}
\vskip.2truecm
\noindent
In the last year there has been a resurging interest in supergravity
theories, especially in  eleven-dimensions [1] and ten-dimensions [2]
in connection
with  duality proporties of superstring theories [3]. There are,
however, supergravity theories which do not correspond to superstring
theories, notably eleven-dimensional supergravity and
ten-dimensional $N=1$ supergravity formulated with
a six-form [4] rather than a two-form. The two formulations in
ten-dimensions
are dual to each other, even in the presence of super  Yang-Mills
multiplet or higher curvature terms [4,5]. The formulation
of supergravity with  the
six-form has been conjectured to be  the low-energy limit of the
four-brane [6], but since the quantization of membranes
gives continuous spectra [7],
the link at present is not clearly established. Further,  eleven
dimensional supergravity, when compactified on a circle, is thought to
be equivalent to the type II superstring [3], and this in turn
have some compactifications identical to those of the heterotic
string when solitonic modes are taken into account. On the other
hand eleven dimensional supergravity does not admit a dual formulation
and the three-form could not be replaced with a six-form [9], although the
two-brane
theory is conjectured to admit a dual five-brane [6].
The trivial dimensional
reduction of eleven-dimensional supergravity to ten-dimensiosn
gives type IIA non-chiral $N=2$ supergravity [11], and when
truncated to
$N=1$ supergravity is known to have a dual formulation [2,4]. What prevents
the eleven-dimensional theory from admitting a dual form is the
presence of a Chern-Simons term $ \int A_3 \wedge dA_3 \wedge dA_3 $ where
$A_3$ is a three form. As $A_3$ appears explicitely in the
action  a duality transformation is not possible, because
the action cannot be expressed solely in terms of the field
strength.

The purpose of this note is to show that type IIA $N=2$ supergravity
in ten-dimensions, although obtained from the eleven dimensional
theory by trivial dimensioanl reduction, admits more than one
dualisation. As the three form in eleven dimensions is reduced
to a three-form and a two-form in ten-dimensions, the Chern-Simons
term can always be rewritten in terms of either the field strength of the
two-form or three-form. Further, a one-form which originates from the
eleven dimensional metric, mixes with the two and three forms, in
such a way that one of the one-form, two-form, or three-form appears
only through its field strength . This will allow us to pass
to the dual versions in ten-dimensions, provided one shows that
the supersymmetry invariance continues to hold for the dual action.
The plan of this paper is as follows.
In section two  we start from the type IIA supergravity in
ten dimensions, and show the modifications needed to allow
for the duality transformations to be performed. In sections
three, four and five we perform these transformations to obtain
three other dual versions of the theory. In section 6 we
indicate how to obtain another dual formulation and the
connection between this family of theories.
%Section 2

{\bf \noindent 2. $N=2$ supergravity in ten-dimensions: (1,2,3) }
\vskip .2truecm
\noindent
Non-chiral $N=2$ supergravity in ten dimensions was obtained [9]
by trivialy reducing the eleven dimensional theory [1].
 The action is expressed in
terms of the bosonic fields $A_{\mu\nu}$ (or $A_2$),
$A_{\mu\nu\rho}$ (or $A_3$), $B_{\mu}$ (or $B$),
$\phi $ and the vielbein $e_{\mu}^a $.
Because of the presence of the one-form, two-form and three-form
we will denote this formulation by (1,2,3).
  The fermionic fields are
the gravitino $\psi_{\mu}$ and the spinor $\lambda $ both
of which are Majorana spinors. The action is given by [9]
$$
\eqalign{
I &=\int d^{10}x e\Bigl( -{1\over {4 \kappa^2}} R(\omega (e) )
-{i\over 2} \overline{\psi}_{\mu} \Gamma^{\mu\nu\rho} D_{\nu} \psi_{\rho}
-{1\over {48}} e^{\kappa \phi} F_{\mu\nu\rho\sigma}^{\prime}
F^{\prime\mu\nu\rho\sigma} \cr
& \qquad +{1\over {12}} e^{-2\kappa \phi} F_{\mu\nu\rho}F^{\mu\nu\rho}
-{1\over 4} e^{3\kappa \phi} G_{\mu\nu}G^{\mu\nu} +{1\over 2}\partial_{\mu}
\phi \partial^{\mu}\phi \cr
& \qquad +{i\over 2} \overline{\lambda} \Gamma^{\mu} D_\mu \lambda
-{i\kappa \over {\sqrt 2}} \overline {\lambda} \Gamma^{11} \Gamma^{\mu}
\Gamma^{\nu}\psi_{\mu} \partial_{\nu}\phi \cr
&\qquad +{\kappa \over {8(12)^2}}e^{-1}\epsilon^{\mu_1 \cdots \mu_{10}}
F_{\mu_1 \cdots \mu_4}F_{\mu_5 \cdots \mu_8} A_{\mu_9\mu_{10}} \cr
& \qquad +{\kappa \over {96}} e^{\kappa \phi\over 2 } \bigl(
\overline {\psi}_{\mu} \Gamma^{\mu\nu\alpha\beta\gamma\delta}\psi_{\nu}
+12 \overline{\psi}^{\alpha}\Gamma^{\beta\gamma}\psi^{\delta}
+{1\over \sqrt 2}\overline{\lambda} \Gamma^{\mu}\Gamma^{\alpha\beta
\gamma\delta}\psi_{\mu} +{3\over 4} \overline{\lambda} \Gamma^{\alpha
\beta\gamma\delta}\lambda \bigr) F_{\alpha\beta\gamma\delta}' \cr
&\qquad -{\kappa\over {24}}e^{-\kappa\phi\over 3}
\bigl( \overline{\psi}_{\mu}
\Gamma^{11} \Gamma^{\mu\nu\alpha\beta\gamma}\psi_{\nu} -6
\overline {\psi}^{\alpha} \Gamma^{11} \Gamma^{\beta}\psi^{\gamma}
-\sqrt 2 \overline {\lambda} \Gamma^{\mu} \Gamma^{\alpha\beta\gamma}
\psi_{\mu}\bigr) F_{\alpha\beta\gamma} \cr
&\qquad -{i\kappa\over 8} e^{3\kappa\phi\over 2}\bigl(
\overline{\psi}_{\mu}\Gamma^{11} \Gamma^{\mu\nu\alpha\beta}\psi_{\nu}
+2\overline{\psi}^{\alpha}\Gamma^{11}\psi^{\beta} +{3\over \sqrt 2}
\overline{\lambda} \Gamma^{\mu}\Gamma^{\alpha\beta}\psi_{\mu}
+{5\over 4} \overline{\lambda} \Gamma^{11} \Gamma^{\alpha\beta}\lambda
\bigr) G_{\alpha\beta} \cr
&\qquad +{\rm quartic \ fermionic \ terms}\Bigr) \cr}\eqno(2.1)
$$
where $G_{\mu\nu}$, $F_{\mu\nu\rho}$ and $F_{\mu\nu\rho\sigma}$
are field strengths of $B_{\mu}$, $A_{\mu\nu}$ and $A_{\mu\nu\rho}$
respectively. Because of the eleven dimensional origin of this
theory one has the modified field strength
$ F_{\mu\nu\rho\sigma}^{\prime} $
where
$$\eqalign{
G_{\mu\nu} &=2\partial_{[\mu}B_{\nu ]} \cr
F_{\mu\nu\rho} &=3\partial_{[\mu}A_{\nu\rho ]} \cr
F_{\mu\nu\rho\sigma}^{\prime} &=4\bigl( \partial_{[\mu}A_{\nu\rho\sigma ]}
+2B_{[\mu}F_{\nu\rho\sigma ]}\bigr) \cr}\eqno(2.2)
$$
As can be seen by compactifying the eleven-dimensional theory
working in a flat frame [4,11], we can write the field strength
 $F'$ in terms
of a modified potential $A_3'$, where
$$\eqalign{
A_{\mu\nu\rho}^{\prime} &= A_{\mu\nu\rho}-6B_{[\mu}A_{\nu\rho ]} \cr
F_{\mu\nu\rho\sigma}^{\prime} &= 4 \bigl( \partial_{[\mu}A_{\nu\rho\sigma ]}'
+3 G_{[\mu\nu}A_{\rho\sigma ]} \bigr) \cr}\eqno(2.3)
$$
These identities will play a vital role in allowing for duality
transformations. The supersymmetry transformations are given by
$$\eqalign{
\delta e_{\mu}^a &= -i\overline{\epsilon} \Gamma^a \psi_{\mu} \cr
\delta \psi_{\mu} &= D_{\mu}(\omega ) -{1\over 32} e^{3\kappa\phi
\over 2}\bigl(\Gamma_{\mu}^{\ \nu\rho} -14 \delta_{\mu}^{\nu}\Gamma^{\rho}
\bigr) \Gamma^{11} \epsilon G_{\nu\rho} \cr
&\qquad +{i\over 48} e^{-\kappa\phi} \bigl( \Gamma_{\mu}^{\ \nu\rho\sigma}
-9\delta_{\mu}^{\nu}\Gamma^{\rho\sigma}\bigr) \Gamma^{11} \epsilon
F_{\nu\rho\sigma} \cr
&\qquad +{i\over 128} e^{\kappa\phi\over 2} \bigl( \Gamma_{\mu}^
{\ \nu\rho\sigma\tau} -{20\over 3} \delta_{\mu}^{\nu} \Gamma^{\rho
\sigma\tau} \bigr) \epsilon F_{\nu\rho\sigma\tau}' +\cdots \cr
\delta B_{\mu} &= {i\over 2}e^{-{3\over 2}\kappa\phi}\bigl(
\overline{\psi}_{\mu} \Gamma^ {11} \epsilon -{\sqrt 2 \over 4}
\overline{\lambda} \Gamma_{\mu} \epsilon \bigr) \cr
\delta A_{\mu\nu} &=e^{\kappa\phi} \bigl( \overline{\psi}_{[\mu}
\Gamma_{\nu ]}\Gamma^{11} \epsilon -{1\over 2\sqrt 2} \overline{\lambda}
\Gamma_{\mu\nu} \epsilon \bigr) \cr
\delta A_{\mu\nu\rho} &= -{3\over 2}e^{-{\kappa\phi\over 2}}\bigl(
\overline{\psi}_{[\mu} \Gamma_{\nu\rho ]}\epsilon -{1\over
6\sqrt 2}\overline{\lambda} \Gamma^{11} \Gamma_{\mu\nu\rho}\epsilon
\bigr) \cr
&\qquad +6e^{\kappa\phi}
B_{[\mu}\bigl(\overline{\psi}_{\nu}\Gamma_{\rho ]}\Gamma^{11} \epsilon
-{1\over 2\sqrt 2} \overline{\lambda} \Gamma_{\nu\rho ]} \bigr) \cr
\delta \lambda &= {1\over \sqrt 2} D_{\mu}\phi (\Gamma^{\mu}
\Gamma^{11} \epsilon ) +{3\over 8\sqrt 2} e^{3\kappa\phi\over 2}
\Gamma^{\mu\nu}\epsilon G_{\mu\nu} \cr
&\qquad +{i\over 12\sqrt 2} e^{-\kappa\phi} \Gamma^{\mu\nu\rho}
\epsilon F_{\mu\nu\rho} +\cdots \cr
\delta \phi &= {i\over \sqrt 2} \overline{\lambda} \Gamma^{11} \epsilon
\cr}\eqno(2.4)
$$

Another important piece is the Chern-Simons term which can be
written in terms of differential forms as $ \int A_2 \wedge
dA_3 \wedge dA_3 $ where
$A_2$ and $A_3$ stand for the two- and three-forms:
$A_2=A_{\mu\nu}dx^{\mu}\wedge dx^{\nu}$, and
$A_3=A_{\mu\nu\rho}dx^{\mu}\wedge dx^{\nu}\wedge dx^{\rho}$. This
can be reexpressed in such a way that
 $A_{\mu\nu}$ appears only through  its field strength $F_{\mu\nu\rho}$.
We derive this by using
$$
A_2 \wedge dA_3 \wedge dA_3 =d(A_2\wedge A_3\wedge dA_3)
 -dA_2 \wedge A_3 \wedge dA_3 \eqno(2.5)
$$
and discarding the surface term after integration.
Next, although the field $B_{\mu}$ does not appear in the
Chern-Simons term, it appears explicitely in the
field strength $F_{\mu\nu\rho\sigma}'$ in eq (2.2). If
equation (2.3) is used instead of (2.2), then $B_{\mu}$ appears
only through its field strength $G_{\mu\nu}$ but then the
Chern-Simons term must  be expressed in terms of
$A_{\mu\nu\rho}'$. It is not difficult to show that
$$\eqalign{
A_2 \wedge dA_3 \wedge dA_3 &= A_2 \wedge d A_3' \wedge dA_3'
 +6 A_2\wedge A_2 \wedge dB \wedge dA_3' \cr
&\qquad +12 A_2\wedge A_2\wedge A_2 \wedge dB \wedge dB \cr
&\qquad +6 d\bigl(A_2\wedge A_2\wedge B \wedge (dA_3' +4A_2\wedge dB )\bigr)
 \cr} \eqno(2.6)
$$
Discarding the surface term, we see that the action (2.1) is
expressible in terms of $A_2$, $dA_3'$ and $dB$. From all of
these considerations it is very suggestive that we can apply
 duality transformations to the following fields $(A_6, A_2)$,
 or $(A_3, A_2)$ or $(B, A_7)$. We now consider these
transformations one at a time.

%section 3

{\bf\noindent 3. A dual theory with a six-form $A_6$: (1,6,3) }
\vskip.2truecm
\noindent
To obtain the dual theory where the two-form  is replaced
with a six-form, we add to the action (2.1) the term
$$
{1\over {3!6!}}\int A_6 \wedge dF_3  \eqno(3.1)
$$
where $A_6 = A_{\mu_1\cdots \mu_6}dx^{\mu_1}\wedge \cdots \wedge
dx^{\mu_6} $
is a six-form and $F_3$ is a three-form, $F_3 =F_{\mu\nu\rho}
dx^{\mu}\wedge dx^{\nu}\wedge dx^{\rho} $,
which in (2.1)
is not assumed now to be a field strength. The equation
of motion of $A_6$ forces  $F_3$, locally, to be $dA_2$.
Integrating by parts and discarding the surface term,
eq (3.1) can be rewritten in the form
$$
{1\over {3!6!}} \int F_3 \wedge dA_6   \eqno (3.2)
$$
Since $F_3$ appears in the action (2.1) and (3.2) at most
quadratically, we can perform the $F_3$ gaussian integration to
obtain the dual version as a function of $A_6$.
Therefore, the action in the form (2.1) plus (3.1) can give either one
of the two dual actions, depending on what is integrated
first, $A_6$ or $F_3$.  The
supersymmetry transformations of  the combined action can be found
as follows [10]. The supersymmetry transformations of $F_3$ are
taken to be identical to those of $d\delta A_2 $ as given
in eq (2.2) (without identifying $F_3$ with $dA_2$ ), then
the action (2.1) will be invariant except for one term proportional
to $dF_3$ which does not vanish now because the Bianchi
identity is no  longer available. The non-invariant term will be cancelled
by the transformation of the new term (3.1) which is also
proportional to $\int \delta A_6 \wedge dF_3 $ . This determines
$ \delta A_6 $ to be given by
$$
\delta A_{\mu_1 \cdots \mu_6} =-3i e^{-\kappa \phi }
\bigl( \overline {\epsilon} \Gamma_{[\mu_1 \cdots \mu_5}\psi_{\mu_6]}
-{i\over {6\sqrt 2}} \overline{\epsilon} \Gamma_{\mu_1 \cdots \mu_6}
\Gamma^{11} \lambda \bigr) \eqno(3.3)
$$
and explicitely shows that the action (2.1) plus (3.1) admits a
duality transformation between the two-form and the six-form.
The duality transformation is at the level of the action and not
only the equations of motion. As
the field $F_{\mu\nu\rho}$ appears at most quadratically, doing
the gaussian integration for $F_{\mu\nu\rho}$, or solving its equation
of motion and substituting back into the action, are equivalent.
The equation of motion gives
$$
M_{\alpha\beta\gamma}^{\mu\nu\rho}F_{\mu\nu\rho}=X_{\alpha\beta\gamma}
\eqno(3.4)
$$
where the tensors $M_{\alpha\beta\gamma}^{\mu\nu\rho}$ and
$X_{\alpha\beta\gamma}$ are given by
$$\eqalignno{
M_{\alpha\beta\gamma}^{\mu\nu\rho} &= \Bigl( {1\over {3!6}}
e^{-2\kappa \phi}(1-e^{3\kappa\phi}B_{\sigma}B^{\sigma})
\delta_{\alpha\beta\gamma}^{\mu\nu\rho} +{1\over 4}e^{\kappa \phi}
B_{[\alpha}\delta_{\beta\gamma}^{[\nu\rho}B^{\mu ]} \Bigr) & (3.5)\cr}
$$
$$\eqalignno{
X_{\alpha\beta\gamma} &=-{1\over {216}} \epsilon_{\alpha\beta\gamma}^
{\ \ \ \ \ \mu_1 \cdots \mu_7} \bigl( {1\over {7!}} F_{\mu_1\cdots \mu_7}
+A_{\mu_1\mu_2\mu_3}\partial_{\mu_4}A_{\mu_5\mu_6\mu_7} \bigr) \cr
&\qquad +{\kappa \over {24}} e^{-{1\over 3}\kappa \phi }\bigl(
\overline{\psi}_{\mu} \Gamma^{11} \Gamma^{\mu\nu}_{\ \ \alpha\beta\gamma}
\psi_{\nu}
-6\overline{\psi}_{[\alpha}\Gamma^{11}\Gamma_{\beta}\psi_{\gamma ]}
-\sqrt 2 \overline{\lambda} \Gamma^{\mu}\Gamma_{\alpha\beta\gamma}
\psi_{\mu}\bigr) \cr
&\qquad -{\kappa \over {12}}e^{{1\over 2}\kappa\phi }\bigl(
\overline{\psi}_{\mu}\Gamma^{\mu\nu\rho}_{\ \ \  \alpha\beta\gamma} \psi_{\nu}
+12\overline{\psi}^{\rho}\Gamma_{[\alpha \beta}\psi_{\gamma ]} \cr
&\qquad \qquad \qquad +{1\over {\sqrt 2}} \overline{\lambda} \Gamma^{\mu}
\Gamma^{\rho}_{\ \alpha\beta\gamma}\psi_{\mu} +{3\over 4} \overline{\lambda}
\Gamma^{\rho}_{\ \alpha\beta\gamma}\lambda \bigr) B_{\rho} &(3.6)\cr}
$$
and we have denoted $F_{\mu_1\cdots\mu_7}=7\partial_{[\mu_1}A_{\mu_2
\cdots\mu_7 ]} $.

Solving equation (3.4) for $F_{\mu\nu\rho}$ gives
$$
F_{\mu\nu\rho}=M_{\ \ \ \mu\nu\rho}^{-1\alpha\beta\gamma}X_{\alpha\beta\gamma}
\eqno(3.7)
$$
where the tensor $ M_{\ \ \ \mu\nu\rho}^{-1\alpha\beta\gamma}$ is the inverse
of $M_{\alpha\beta\gamma}^{\mu\nu\rho}$:
$$M_{\ \ \ \mu\nu\rho}^{-1\alpha\beta\gamma}M_{\alpha\beta\gamma}^{\kappa
\lambda\sigma} ={1\over {3!}}\delta_{\mu\nu\rho}^{\kappa\lambda\sigma}
\eqno(3.8)
$$
The explicit form of $M^{-1}$ is
$$
M_{\ \ \ \mu\nu\rho}^{-1 \alpha\beta\gamma}={6
e^{2\kappa\phi}\over {1-e^{3\kappa\phi}B_{\sigma}B^{\sigma}}}\Bigl(
{1\over {3!}}\delta_{\alpha\beta\gamma}^{\mu\nu\rho}-{3\over
2}e^{\kappa \phi} B_{[\mu} \delta_{\nu\rho
]}^{[\alpha\beta}B^{\gamma ]}\Bigr) \eqno(3.9)
$$
Therefore to obtain the dual action from (2.1) plus (3.1), we discard
all the $F_{\mu\nu\rho}$ contributions and replace them with
$$
-{1\over 2} X_{\alpha\beta\gamma} M_{\ \ \ \mu\nu\rho}^{-1 \alpha\beta\gamma}
X_{\mu\nu\rho}   \eqno(3.11)
$$

%section 4
{\bf\noindent 4. The dual action with a five-form: (1,2,5)}
\vskip.2truecm
\noindent
To find the $N=2$ supergravity action where the three-form
is replaced with a five-form we proceed as before. First,
we write the action (2.1) in such a way that  the three-form
appears only through its field strength. We
use eq (2.3) for $F_{\mu\nu\rho\sigma}'$, and
write it as $F_{\mu\nu\rho\sigma}+12 G_{[\mu\nu}A_{\rho\sigma ]}$.
Then we assume that
$F_{\mu\nu\rho\sigma }$ is an independent field
and not the field strength of $A_{\mu\nu\rho}'$, and
add the following term to the action:
$$
{1 \over {4!5!}}\int A_5 \wedge dF_4  \eqno(4.1)
$$
where $A_5=A_{\mu_1\cdots \mu_5}dx^{{\mu}_1}\wedge \cdots
\wedge dx^{{\mu}_5} $. The
$A_5$ equation implies, locally, that $F_4=dA_3' $ and this gives again
the action (2.1). If, however, we integrate eq (4.1) by parts,
and then do the gaussian integration of $F_{\mu\nu\rho\sigma}$
we will be left with an action in terms of the dual field
strength $F_{\mu_1\cdots\mu_6}$. To restore the supersymmetry
invariance after adding (4.1) to the action (2.1) we assume
that $\delta F_{\mu\nu\rho\sigma} =4\partial_{[\mu} \delta
A_{\nu\rho\sigma ]}$, then the extra terms that spoil the invariance
of the action (2.1) are cancelled by those arising from the
non-invariance of the term (4.1). This
is achieved by taking
$$
\delta A_{\mu_1\cdots\mu_5}={5\over 2}i e^{{1\over 2}\kappa \phi }
\overline{\epsilon} \Gamma^{11}
\Gamma_{[\mu_1\cdots\mu_4}\psi_{\mu_5 ]} \eqno(4.2)
$$
The sum of the actions (2.1) and (4.1)  gives both
dual actions depending on the order of integration and is
invariant under the new supersymmetry transformations.

The gaussian integration of $F_{\mu\nu\rho\sigma}$ gives
$$
{1\over 2}  X_{\mu\nu\rho\sigma}M_{\ \ \ \alpha\beta\gamma\delta}^{-1
\mu\nu\rho\sigma} X^{\alpha\beta\gamma\delta} \eqno(4.3)
$$
where $X_{\mu\nu\rho\sigma}$ is defined by
$$\eqalign{
X_{\mu\nu\rho\sigma} &= {\kappa \over {4!}} \epsilon_{\mu\nu\rho\sigma}
^{\ \ \ \ \ \mu_1\cdots\mu_6}\bigl( {1\over 6!} F_{\mu_1\cdots\mu_6}
+{1\over 16} A_{\mu_1\mu_2}A_{\mu_3\mu_4}G_{\mu_5\mu_6} \bigr) \cr
&\qquad +{\kappa \over 96}e^{{1\over 2}\kappa \phi }\bigl( \overline
{\psi}_{\rho}\Gamma^{\alpha\beta}_{\ \ \mu\nu\rho\sigma}\psi_{\beta}
+12 \overline{\psi}_{[\mu} \Gamma_{\nu\rho}\psi_{\sigma ]}
+{1\over {\sqrt 2}} \overline{\lambda}\Gamma^{\alpha}\Gamma_{
\mu\nu\rho\sigma}\psi_{\alpha} +{3\over 4} \overline{\lambda}
\Gamma_{\mu\nu\rho\sigma}\lambda\bigr) \cr
&\qquad -{1\over 2} e^{\kappa\phi} G_{[\mu\nu}A_{\rho\sigma ]}\cr}\eqno(4.4)
$$
and the matrix $M^{-1}$ is the inverse of
$$
M_{\mu\nu\rho\sigma}^{\alpha\beta\gamma\delta}=({1\over 4!})^2
\Bigl( e^{\kappa\phi}\delta_{\mu\nu\rho\sigma}^{\alpha\beta\gamma
\delta} -\kappa \epsilon_{\mu\nu\rho\sigma}^{\ \ \ \ \alpha
\beta\gamma\delta\lambda\tau}A_{\lambda\tau} \Bigr) \eqno(4.5)
$$
defined by
$$
M_{\ \ \ \mu\nu\rho\sigma}^{-1\alpha\beta\gamma\delta}
M_{\alpha\beta\gamma\delta}^{\kappa\lambda\tau\eta}
={1\over {4!}}\delta_{\mu\nu\rho\sigma}^{\kappa\lambda\tau\eta}
\eqno(4.6)
$$
The explicit expression of $M^{-1}$ is too long to give here.
The field strength $F_6$ is given by
$$
F_{\mu_1\cdots\mu_6}=6\partial_{[\mu_1} A_{\mu_2\cdots\mu_6 ]} \eqno(4.7)
$$

Therefore, to obtain the dual action we discard all the terms
containing $F_{\mu\nu\rho\sigma}$ and replace them with  (4.3).
This completes the derivation of the dual action where the
three-form is replaced by the five-form.
\vskip.2truecm
%section 5
{\bf \noindent 5. The dual action with a seven-form: (7,2,3) }
\vskip.2truecm
\noindent
As we have seen in section 2, there exists the possibility of
writing the $N=2$ supergravity action IIA in such a way that
the one-form $B$ appears only through its field strength.
This required a redefinition of the three-form. The procedure
of obtaining the action where the one-form is replaced with
the seven-form is the same as before. We first manipulate
the action (2.1) so that the field $B_{\mu}$ appears only
through its field strength $G_{\mu\nu}$ then we assume that
$G_{\mu\nu}$ is an independent field and add a term to the
action (2.1) of the form:
$$
{1\over{2!7!}} \int A_7 \wedge d G  \eqno(5.1)
$$
where we have defined the seven-form
$A_7=A_{\mu_1\cdots\mu_7}dx^{\mu_1}\wedge \cdots \wedge dx^{\mu_7}$
Integrating the $A_7$ field out implies the constraint
$dG=0$  whose solution , locally, is
$G_{\mu\nu}=2\partial_{[\mu}B_{\nu ]}$ and this takes us back
to the action (2.1). Integrating the action (5.1) by parts
and discarding the surface term we obtain
$$
{{1\over {2!7!}}} \int dA_7 \wedge G  \eqno(5.2)
$$
The field $F_{\mu\nu\rho\sigma}'$ in the action (2.1) is
taken to be of the form (2.3) and the Chern-Simons term is
rearranged to be given by (2.5). Then the full action is
at most quadratic in $G_{\mu\nu}$ and the gaussian
integration can be performed. This will give the dual
action expressed in terms of the field strength of $A_7$.
The non-invariance of (2.1) under the supersymmetry transformations
due to the removal of the identificaiton $G=dB$ is cancelled
by the varriation of (5.1) provided one identifies the
varriation of $G$ with
$$
\delta G_{\mu\nu}=2\partial_{[\mu}\delta B_{\nu ]}  \eqno(5.3)
$$
and the varriation of $A_7$ with
$$\delta A_{\mu_1\cdots\mu_7} =e^{{3\over 2}\kappa \phi}\bigl(
-{7\over 2} \overline{\epsilon}\Gamma_{[\mu_1\cdots\mu_6}\psi_{\mu_7 ]}
+{\sqrt 2\over 8} \overline{\epsilon} \Gamma_{\mu_1\cdots\mu_7}
\Gamma^{11}\lambda \bigr) \eqno(5.4)
$$
The gaussian integration of $G_{\mu\nu}$ gives
$$
-{1\over 4} X_{\mu\nu}M_{\ \ \ \alpha\beta}^{-1\mu\nu}X^{\alpha\beta}
\eqno(5.5)
$$
where $X_{\mu\nu}$ is defined by
$$\eqalign{
X_{\mu\nu} &=-{1\over 8}e^{\kappa\phi}A^{\rho\sigma}\bigl(
4\partial_{[\mu} A_{\nu\rho\sigma ]}'\bigr) \cr
&\qquad +{3\kappa\over {16}} \bigl( \overline{\psi}_{\alpha}
\Gamma^{\alpha\beta}_{\ \ \mu\nu\rho\sigma}\psi_{\beta}
+12\overline{\psi}_{[\mu}\Gamma_{\nu\rho}\psi_{\sigma ]}
+{1\over {\sqrt 2}}\overline{\lambda} \Gamma^{\alpha}
\Gamma_{\mu\nu\rho\sigma}\psi_{\alpha} +{3\over 4}
\overline{\lambda}\Gamma_{\mu\nu\rho\sigma}\lambda \bigr)
A^{\rho\sigma} \cr
&\qquad -{i\kappa \over 8}e^{{3\over 2}\kappa \phi}\bigl(
\overline{\psi}_{\alpha}\Gamma^{11}\Gamma^{\alpha\beta}_{\ \ \mu\nu}
\psi_{\beta} +2\overline{\psi}_{[\mu}\Gamma^{11} \psi_{\nu ]}
+{3\over 2} \overline{\lambda} \Gamma^{\alpha}\Gamma_{\mu\nu}\psi_{\alpha}
+{5\over 4}\overline{\lambda} \Gamma^{11}\Gamma_{\mu\nu}\lambda \bigr)\cr
&\qquad +\epsilon_{\mu\nu}^{\ \ \ \mu_1\cdots\mu_8} \Bigl(
{1\over{2!7!}} \partial_{\mu_1}A_{\mu_2\cdots\mu_8}
-{\kappa \over 192}A_{\mu_1\mu_2}\cdots A_{\mu_7\mu_8} \Bigr) \cr}\eqno(5.6)
$$
and the tensor $M_{\ \ \ \mu\nu}^{-1\alpha\beta}$ is the inverse
of
$$\eqalign{
M_{\mu\nu}^{\alpha\beta} &= {1\over 4}e^{3\kappa \phi}\bigl(
1+{3\over
2}A_{\rho\sigma}A^{\rho\sigma}\bigr)\delta_{\mu\nu}^{\alpha\beta}
+A_{\mu\nu}A^{\alpha\beta} \cr
&\qquad -4\delta_{[\mu}^{[\alpha}A_{\nu ]\rho}A^{\beta ]\rho}
-{\kappa\over 96}\epsilon_{\ \ \ \mu\nu\mu_1\cdots\mu_6}^{\alpha
\beta}A_{\mu_1\mu_2}A_{\mu_3\mu_4}A_{\mu_5\mu_6} \cr}\eqno(5.7)
$$
The inverse of $M$ is defined by:
$$
M_{\ \ \ \mu\nu}^{-1\alpha\beta}M_{\alpha\beta}^{\rho\sigma}={1\over {2!}}
\delta_{\mu\nu}^{\rho\sigma} \eqno(5.8)
$$
but again the explicit expression is too long to give here.
Finally, $G_{\mu\nu}$ is related to its dual by the relation
$$
G_{\mu\nu}=M_{\ \ \ \mu\nu}^{-1\alpha\beta}X_{\alpha\beta}\eqno(5.8)
$$
The dual action is obtained by discarding all the $G_{\mu\nu}$
contributions in (2.1) plus (5.2) and replacing them with (5.5).
This completes the derivation of the dual action where the one-form
is replaced with a seven-form.

%section6
{\bf \noindent 6. Conclusion: connections betwee the different formulations }
\vskip.2truecm
\noindent
In this letter we have shown that the original formulation
of $N=2$ supergravity type IIA given in terms of a one-form, a
two-form
and a three-form  (we denote this by (1,2,3)), admits three
other dual formulations. In the first, the two-form is replaced
with a six-form giving rise to a formulation in terms of a one-form,
a six-form and a three-form (denoted by (1,6,3)). In the second the
three-form is replaced with a five-form giving rise to
a formulation in terms of (1,2,5) forms. Finally, in the
third the one-form is replaced with a seven-form
giving rise to the (7,2,3) formulation.
It is easy to see that the (1,2,5) formulation depends on
the three-form through its field strength suggesting that
it is possible to find a duality transformation  that takes the
one-form
to a sevem-form. This will give  the (7,2,5) formulation.
 This can also be reached by performing a duality transformation
on the three-form in the (7,2,3) formulation
as it appears only through its field strength.
This also implies that the (7,2,5) formulation can be
reached by applying a double duality transformation to
the one-form and three-form simultaneously. If we arrange
the (1,2,3), (7,2,3), (7,2,5) and (1,25) formulations
at the corners of a square in a clockwise fashion, then
all adjacent vertices could be transformed to each other
by a simple duality transformation, and the opposite edges by
a double duality transformation. But it seems that the (1,6,3)
formulation can only be connected to the (1,2,3) formulation
as it depends on the one-form and three-form explicitely.
It will be interesting to understand
the relation of these field theories and their duality
proporties to those of extended objects.
\vskip0.3truecm
{\bf\noindent Acknowledgments}\hfill\break
I would like to thank the Institute of Theoretical Physics,
University of California, Santa Barbara,
for hospitality where part of this work
was done.
%\vfill
%\eject
\vskip.4truecm
{\bf \noindent References}
\vskip.2truecm

\item{[1]}E. Cremmer, B. Julia and J. Scherk, {\sl Phys. Lett }
{\bf 76B} (1978) 409.

\item{[2]} A. H. Chamseddine, {\sl Nucl.Phys.} {\bf B185 } (1981)
403;\br
E. A. Bergshoeff, M. de Roo, B. de Witt and P. van Niewenhuizen,
{\sl Nucl. Phys.} {\bf B195} (1982) 97;\br
G. F. Chapline and N. S. Manton, {\sl Phys. Lett} {\bf 120B} (1983)
105.

\item{[3]}C. M. Hull and P. K. Townsend, {\sl Nucl. Phys.}
{\bf B438} (1995) 109;\br
E. Witten, {\sl Nucl. Phys.} {\bf B 443} (1995) 85.

\item{[4]}A. H. Chamseddine, {\sl Phys. Rev.} {\bf D24} (1981)
3065.

\item{[5]}A. H. Chamseddine and P. Nath, {\sl Phys. Rev.}
{\bf D34} (1986) 3769;\br
S. Gates and H. Nishino, {\sl Phys. Lett.} {\bf 173B} (1986) 46;
{\sl Phys. Lett.} {\bf 173B} (1986) 52; {\sl Nucl. Phys.} {\bf B268}
(1986) 532;\br
E. A. Bergshoef and M. de Roo, {\sl Nucl. Phys.} {\bf B328}
(1989) 439.

\item{[6]}M. J. Duff and X. J. Lu, {\sl Nucl. Phys.}
{\bf B354} (1991) 141;{\sl Nucl. Phys.} {\bf B390} (1993) 276;\br
M. J. Duff, R. R. Khuri and X. J. Lu, {\sl Phys. Rept.} {\bf 259}
(1995) 213.

\item{[7]} B. de Witt, M. L\" uscher and H. Nicolai,
{\sl Ncul. Phys.} {\bf B320} (1989) 135.

\item{[8]} H. Nicolai, P. K. Townsend and P. van Niewenhuizen,
{\sl Lett. Nuovo. Cimento} {\bf 30} (1980) 179.

\item{[9]} I. C. Campbell and P. West, {\sl Nucl. Phys.}
{\bf B243} (1984) 112.

\item{[10]} H. Nicolai and P. K. Townsend, {\sl Phys. Lett.}
{\bf 98B} (1981) 257.

\end